\documentclass[aip,reprint,
superscriptaddress,
 amsmath,amssymb,
]{revtex4-2}
\usepackage{graphicx}% Include figure files
\usepackage{dcolumn}% Align table columns on decimal point
\usepackage{bm}% bold math
\usepackage{xcolor}
\usepackage{siunitx}
\begin{document}
\title{Upconversion of Phonon Modes into Microwave Photons in a Lithium Niobate Bulk Acoustic Wave Resonator Coupled to a Microwave Cavity}
%{Coupling of a Macroscopic Lithium Niobate Phonon Cavity to Microwave Photons}
% Force line breaks with \\
\author{S.Parashar$^*$}
\author{W. M. Campbell}%
\author{J.Bourhill}
\author{E.Ivanov}
\author{M.Goryachev}
\author{M.E.Tobar$^*$}
\affiliation{Quantum Technologies and Dark Matter Labs, Department of Physics, University of Western Australia, 35 Stirling Highway, Crawley, WA 6009, Australia.}
\begin{abstract}
The coupling between acoustic vibrations in a lithium niobate bulk acoustic wave resonator and microwave photons of a re-entrant microwave cavity was investigated at a temperature close to 4 K. Coupling was achieved by placing the acoustic resonator in the location of the re-entrant cavity electric field maxima, in a symmetric  ``split-post'' configuration, with a large overlap between the microwave field and the acoustic mode, allowing acoustic modulations of the microwave frequency. We show that the acoustic modes in this setup retain large inherent quality factors of greater than $10^6$.  A maximum optomechanical coupling rate was determined to be $g_0$ = 0.014 mHz, four orders of magnitude larger than previous results obtained using a quartz BAW at 4 K in a similar experimental setup, but using a single post-re-entrant cavity resonator. 
\end{abstract}

\email{23307714@student.uwa.edu.au}
\email{michael.tobar@uwa.edu.au}
\maketitle

Investigating the fundamental interaction between photonic and phononic systems is crucial for advancing various applications within the domain of quantum technology and quantum electrodynamics \citep{aspelmeyer2014cavity,blais2021circuit}. This understanding has led to the emergence of numerous associated predicted phenomena. Resolved sideband cooling \cite{cuthbertson1996parametric,teufel2011sideband}, parametric amplification \cite{wilson2015measurement, kumar2024optomechanically, xiong2018fundamentals}, optomechanically induced transparency/absorption \cite{agarwal2010electromagnetically, kumar2024optomechanically}, and long-range entanglement \cite{yao2024long,palomaki2013entangling}, are just a few of these phenomena that have had important ramifications for testing fundamental quantum limits \citep{schrinski2023macroscopic,bushev2019testing,campbell2023improved,lo2016acoustic,goryachev2018next}, dark matter detection \citep{arvanitaki2016sound,carney2021ultralight,campbell2021searching}, quantum communication, information processing and storage, along with generation and manipulation of quantum states \citep{gokhale2020epitaxial,chu2017quantum, salzenstein2010significant}. Importantly for these systems, highly coherent quantum state readouts and increased coupling rates are integral for parametric detection techniques. This is particularly relevant in the search for gravitons \citep{tobar2023detecting,tobar2024detecting} and in the development of parametric transducers for achieving low noise readouts at microwave frequencies \citep{cuthbertson1996parametric,locke1998parametric}.

Ground state preparation of truly macroscopic (gram-scale) mechanical resonators is a highly sought after experimental demonstration \cite{bushev2019testing, bourhill2015precision, bourhill2020generation}. For one, there still remain questions regarding the quantum-to-classical crossover point which will be revealed by such systems, but in addition, precision readout of such devices may offer insights into quantum gravity and high frequency gravitational wave detection. \citep{goryachev2014gravitational,campbell2023multi, campbell2023improved, goryachev2021rare,singh2017detecting,aggarwal2021challenges}.

\begin{figure}[t]
\centering
\includegraphics[width=0.8\linewidth]{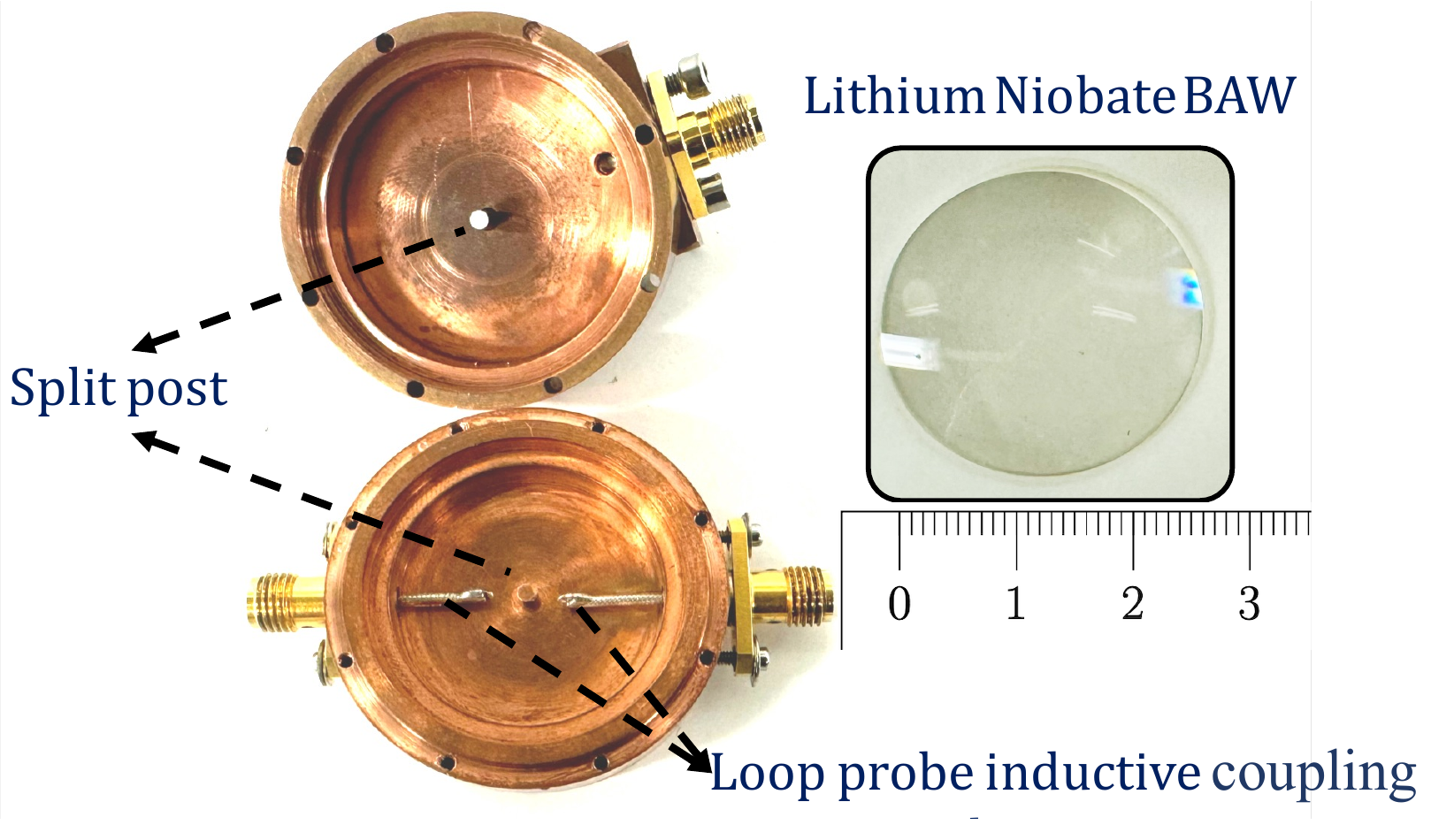}
\caption{The convex Lithium Niobate crystal BAW resonator, along with the two halves of the re-entrant split-post microwave cavity, fabricated from oxygen-free copper.  }
\label{baw-resonator-dimension} 
\end{figure}

Studies on these photon-phonon interactions employ various types of resonant photonic device architectures, such as superconducting circuits \citep{blais2020quantum}, co-planar resonators \cite{manenti2017circuit}, 3-D microwave cavity resonators \citep{kitzman2023free}, and whispering gallery resonators \citep{schliesser2010cavity,salzenstein2010significant,locke2004measurement}, combined with various types of mechanical architectures displaying high coupling rates, such as membrane-in-the-middle systems \citep{kumar2024optomechanically}, trapped ion particles \cite{kotler2017hybrid}, and acoustic-mechanical modes such as bulk acoustic wave (BAW) resonators \citep{bourhill2015precision,schliesser2010cavity,kumar2024optomechanically,bourhill2020generation}. BAW resonators are low-loss mechanical resonators that exhibit long mechanical coherence times due to their specifically engineered convex surface, which helps trap the majority of phonons in the centre of the resonator\citep{goryachev2014gravitational}.

\begin{figure}[t]
\centering
\includegraphics[width=1.0\linewidth]{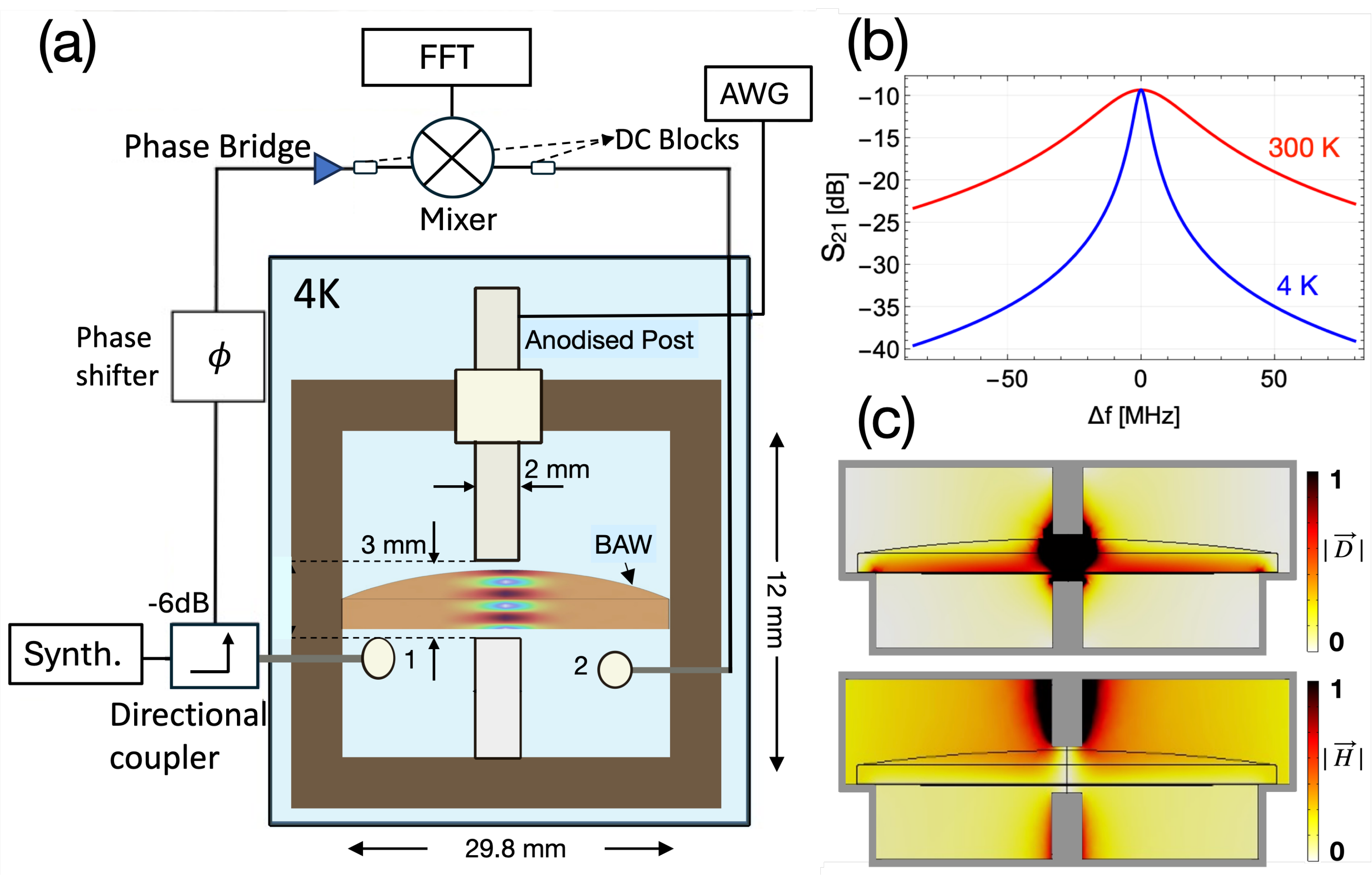}% Here is how to import EPS art
\caption{a) Not to scale microwave-BAW cavity set-up with a frequency discriminator output. The synthesizer was tuned to the microwave cavity and measured at the mixer output with the FFT to detect the upconverted BAW frequency. b) Microwave transmission measurements ($S_{21}$) with a vector network analyser between ports 1 and 2, comparing 300 K with 4K, showing an increase microwave Q-factor. c) To scale electric flux density $|\vec{D}|$  and magnetic field intensity $|\vec{H}|$ density plots inside the split-post microwave cavity. The electric flux density is concentrated in the centre of the BAW in the vicinity of the MHz phonon modes, which are indicted in 2(a).}
\label{Experimental realisation at 4 K}
\end{figure}

In this work, a macroscopic centimetre size-scale lithium niobate (LiNbO$_3$) crystal with its axis cut in the z-direction was manufactured with a diameter of 30 mm, a centre thickness of 2 mm and a convex radius of curvature of 100 mm. The crystal exhibited coherent phonon modes trapped in the centre of the crystal with frequencies of order MHz. Fig.\ref{baw-resonator-dimension} shows a photo of the LiNbO$_3$ crystal along with the two halves of a ``split-post''  microwave cavity resonator\cite{le2013rigorous}, which sandwiches the BAW in the post gap allowing maximum electric field to be applied to the region of maximum phonon trapping as shown in Fig. \ref{Experimental realisation at 4 K}. This means the acoustic phonons will be parametrically upconverted to microwave frequencies. The motivation was drawn from similar experiments conducted with quartz BAW resonators\cite{carvalho2019piezo,bourhill2020generation}. However, the cavity design differs from before as we have used a symmetric design in which the gap is located in the centre of a co-axial cavity. This particular design results in a more highly confined electric field distribution focused into the centre of the BAW, and we term this design the ``split-post re-entrant cavity''. Moreover, the reason for the emerging popularity of LiNbO$_3$ over quartz is that it exhibits a stronger piezoelectric coefficient in addition to a low loss tangent. For example, LiNbO$_3$-BAW has demonstrated a high quality factor $\times$ frequency product, of $\sim$10$^{14}$ Hz at room temperature\cite{campbell2024low}. 

Much smaller LiNbO$_3$ nanostructures have been used previously in various optical applications, such as quantum sensors \cite{marinkovic2021hybrid, cleland2024studying} and optical transducers \cite{arrangoiz2018coupling,satzinger2018quantum,Shao:19,Mirhosseini:2020aa,Jiang:2020aa,wollack2021loss}. Our study investigated LiNbO$_3$ microwave-BAW coupling rates, with the goal of creating strongly coupled coherent microwave readout of large macroscopic phonon modes for fundamental physics applications that require larger masses. Here upconversion opens up possibilities for harnessing effects such as parametric amplification, cooling, and squeezing, while also providing a method for reading out displacement in large non-piezoelectric crystals.

To model the photon-phonon interaction, the dynamics are characterized by the following Hamiltonian\citep{aspelmeyer2014cavity}:
\begin{equation}
        H = \hslash \omega_{\text{c}} a^{\dagger}a +  \hslash \omega_{m} b^{\dagger}b +  \hslash g_0 \left( a^{\dagger}a \right) \left(b^{\dagger} + b \right),
        \label{optomechanical-equation}
\end{equation}
where $a (a^{\dag})$ and $b (b^{\dag})$ are the bosonic annihilation (creation) operators of the microwave and the acoustic resonance, at frequencies $\omega_{\text{c}}$ and $\omega_{\text{m}}$, respectively, and $g_0$ is the optomechnical coupling rate. 

The Hamiltonian in \eqref{optomechanical-equation} describes the interaction between a microwave electric field between the two posts and the acoustic phonon mode of the BAW device, which results in radiation pressure acting on the acoustic modes. This radiation pressure displaces the oscillator (by displacement $x$) and leads to parametric modulation of the microwave cavity fields as shown in \eqref{optomechanical-equation-1}, where the second term on the RHS represents the linear coupling, which can be implemented as a displacement readout\citep{aspelmeyer2014cavity}:
\begin{align}
\omega_{\text{c}}(x) &\approx \omega_{\text{c}} + x\frac{\delta\omega_{\text{c}}}{\delta x} + \ldots. \label{optomechanical-equation-1}
\end{align}
\noindent Furthermore, we can define the following quantities:
\begin{align}
	G = - \frac{\delta\omega_{\text{c}}}{\delta x}  \label{omega_cavity} \ \ \ \text{and} \ \ \ \
	g_0= G x_{\text{zpf}}  ,
\end{align}
where $G$, also known as the frequency pull factor, is measured from the experimental data and helps evaluate $g_0$, which is the photon-phonon coupling rate, also known as the single photon coupling rate.  Additionally, $x_{\text{zpf}} = \sqrt{2\hslash/ \omega_m M_{\text{eff,m}}}$ is the zero-point fluctuation of the mechanical resonance, where $\hslash$ is the reduced Planck's constant, and $M_{\text{eff,m}}$ is the effective mode mass.

Fig. \ref{Experimental realisation at 4 K}(a) illustrates the experimental setup for measurements conducted at 4 K. The experiment utiizes the split post microwave cavity, which functions as a frequency discriminator. This phase bridge detects frequency shifts of the cavity in response to mechanical modes by ensuring that the LO and RF signals are in quadrature, making the system insensitive to amplitude fluctuations. The power splitting ratio is set by a -6dB directional coupler. The signal generator is an analog synthesizer (Keysight E8663D), covering the frequency range of 100 kHz to 9 GHz. Thus the experiment measures the frequency modulation of the BAW resonator up-converted to microwave frequencies. The output of the mixer is observed on the FFT spectrum analyser.

The set-up in Fig. \ref{Experimental realisation at 4 K} also shows the two posts of the microwave cavity, one of which was anodised to provide two electrically isolated posts allowing them to function as electrodes to piezoelectrically excite the BAW mechanical modes with an arbitrary waveform generator (AWG). Since the applied voltage to the electrodes is known, we can estimate the piezoelectric-induced crystal displacement using known LiNbO$_3$ electrostatic material parameters, allowing us to determine the change in displacement $dx$. The measured change in microwave cavity frequency, $d\omega_c$, is then used to determine $G$. 

It is known that for piezoelectric materials, the mechanical displacement $x$ is proportional to the charge on the surface~\cite{aspelmeyer2014cavity}. In our set up, this charge is enhanced due to the split post geometry and is given by, $q = k_mx$, where, $k_m$ is the effective piezoelectric coupling constant defined using the Butterworth-Van Dyke (BWD) model, which can be understood from the BAW-resonator lumped circuit element theory\cite{bible2002crystal}. For a mechanical mode, $m$, relations between piezoelectric constant and the corresponding LCR circuit elements of the BWD model are represented as\cite{campbell2023multi}:
\begin{align}
    M_{\text{eff,m}}= k^2_{m}L_{m};~R_{m}= \frac{k^2_{m}}{C_{m}};~\text{so that}~k^2_{m} = \frac{\omega_m M_{\text{eff,m}}}{Q_{m}R_{m}}.
      \label{4}
\end{align}
Here, $Q_{m}$  is the mode quality factor, $R_{m}$  the motional resistance, and $L_{m}$ and $C_{m}$ the effective inductance and capacitance respectively. Similarly, the current from the equivalent BAW circuit model may be presented in terms of mechanical resonance $\omega_m$ as: 
\begin{equation}
    \label{5}
    I\left(\omega_m\right)= k_m \omega_m x.
\end{equation}
Thus, the values of motional resistance could be measured at 4 K using an impedance analyser. 

The operational cavity mode was the fundamental split-post-mode (Fig. \ref{Experimental realisation at 4 K}c), with a frequency of 6.075 GHz at 4 K.  The loaded quality factor, $Q_L$, improved significantly from 300 to 4 K, to $Q_L=$ 330 to 2500 respectively, due to reduced resistive losses from the copper cavity walls as shown in Fig. \ref{Experimental realisation at 4 K}b. The coupling coefficients were determined from reflection measurements form ports 1 and to, and calculated to be $\beta_1=0.8$ and $\beta_2=0.136$. Thus the unloaded microwave $Q$-factor can thus be determined to be $Q_0=Q_L(1+\beta_1+\beta_2 )=4250$.The increased $Q$-factor helped achieve better discriminator sensitivity for the phase bridge setup. The incident input power to the microwave cavity was 0.01 mW.

The change in the displacement of the mechanical mode is given by the simple relationship between the calculated displacement $\Delta x$ and the change in the output voltage $\Delta V$ from the mixer,  as follows:
\begin{equation}
   \label{2}
    \frac{\Delta V}{\Delta x} = \frac{dV}{d\omega_c}\frac{d\omega_c}{dx},
\end{equation}
where $dV/d\omega_c$ is the discriminator sensitivity (phase-bridge); the ratio at which frequency fluctuations are converted to synchronous voltage fluctuations in the readout, while $G=-d\omega_c/dx$ is the frequency pull factor introduced previously and is evaluated from the voltage response output from the mixer corresponding to the displacement of mechanical resonance. 

From \eqref{optomechanical-equation}, we can also calculate $\frac{d\omega_c}{dx}$ in terms of $x_{\text{zpf}}$:
\begin{equation}
    \label{3}
    \frac{d\omega_c}{dx} = \frac{g_0}{x_{\text{zpf}}}.
\end{equation}
Driving an applied time varying MHz signal across the split-posts, piezoelectrically exciting the BAW, while simultaneously driving the microwave cavity at its resonant frequency, yields a mechanically induced frequency modulation of the microwave pump frequency, producing side bands on the microwave carrier. This is triggered by the induced displacement of the LiNbO$_3$ BAW when tuned to the acoustic modes. Due to the trigonal symmetry of lithium niobate, two-mode families exist. {`}A{'} refers to longitudinal polarization, where bulk displacement occurs in the z-direction, while {`}B{'} denotes a shear wave polarized mode with dominant displacement in the x-y plane. The first subscript refers to the fundamental polarization inside the BAW resonator, which has been studied in \cite{campbell2024low}. More details of the different suffixes in the modes can be found in \cite{goryachev2014gravitational}. In this work, we study the A$_{3,0,0}$ and A$_{5,0,0}$ longitudinal modes, and B$_{5,0,0}$, B$_{7,0,0}$, B$_{9,0,0}$ shear modes. 

\begin{figure}[t]
    \centering
    \includegraphics[width=1.0\linewidth]{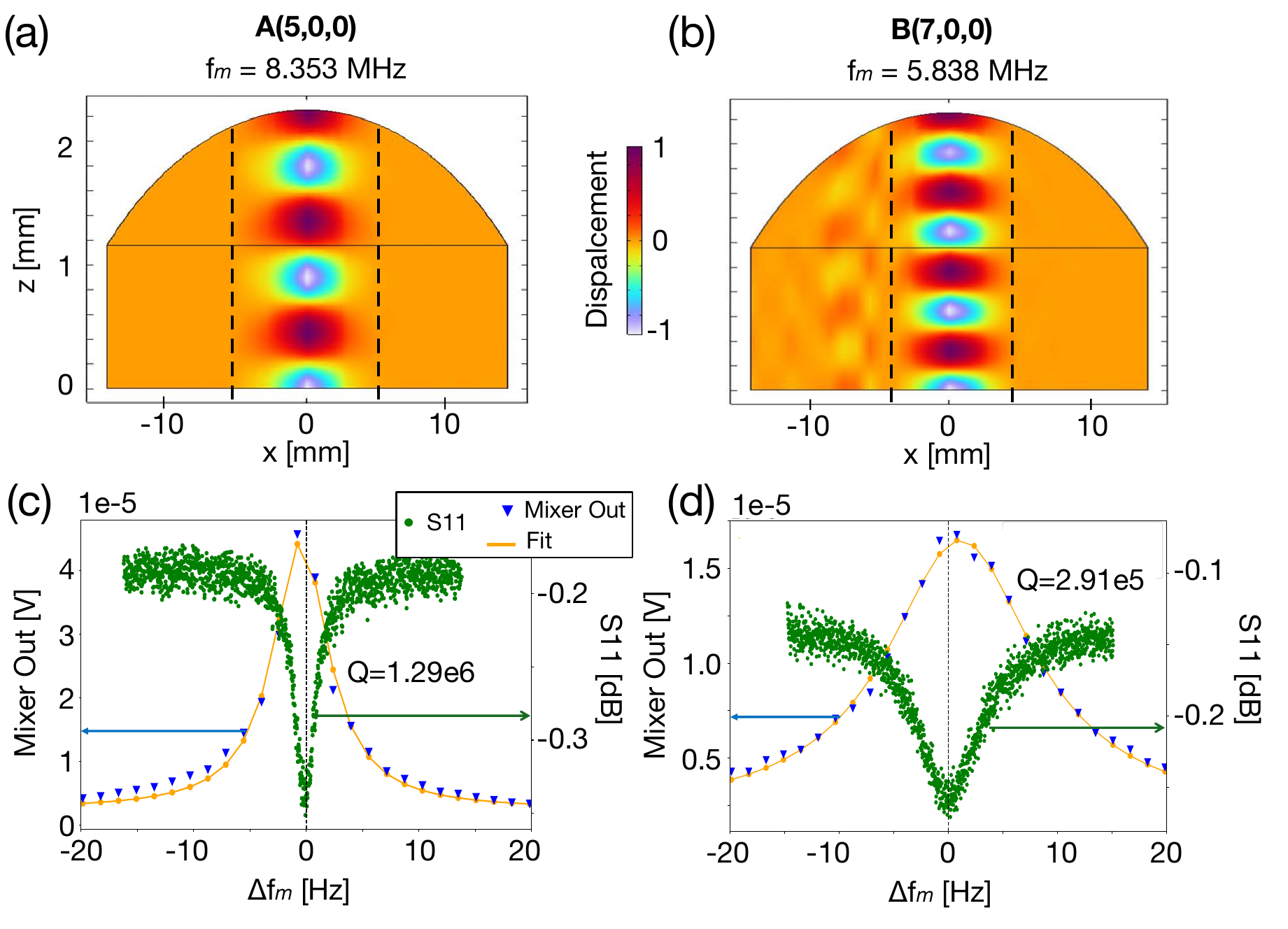}% Here is how to import EPS art
    \caption{FEM simulations of the displacement field of the a) 8.3 MHz longitudinal and b) 5.8 MHz shear modes, with dashed lines indicating the effective mode volumes. c) and d) show the corresponding mixer output voltage in blue and $S_{11}$ reflection measurements in green using an impedance analyser across the split-post. Loaded Q-factors are determined by curve fits, as shown in orange.}
    \label{fig:8.35MHz-5.83MHz} 
    \end{figure}

\begin{figure}[t]
    \centering
    \includegraphics[width=1.0\linewidth]{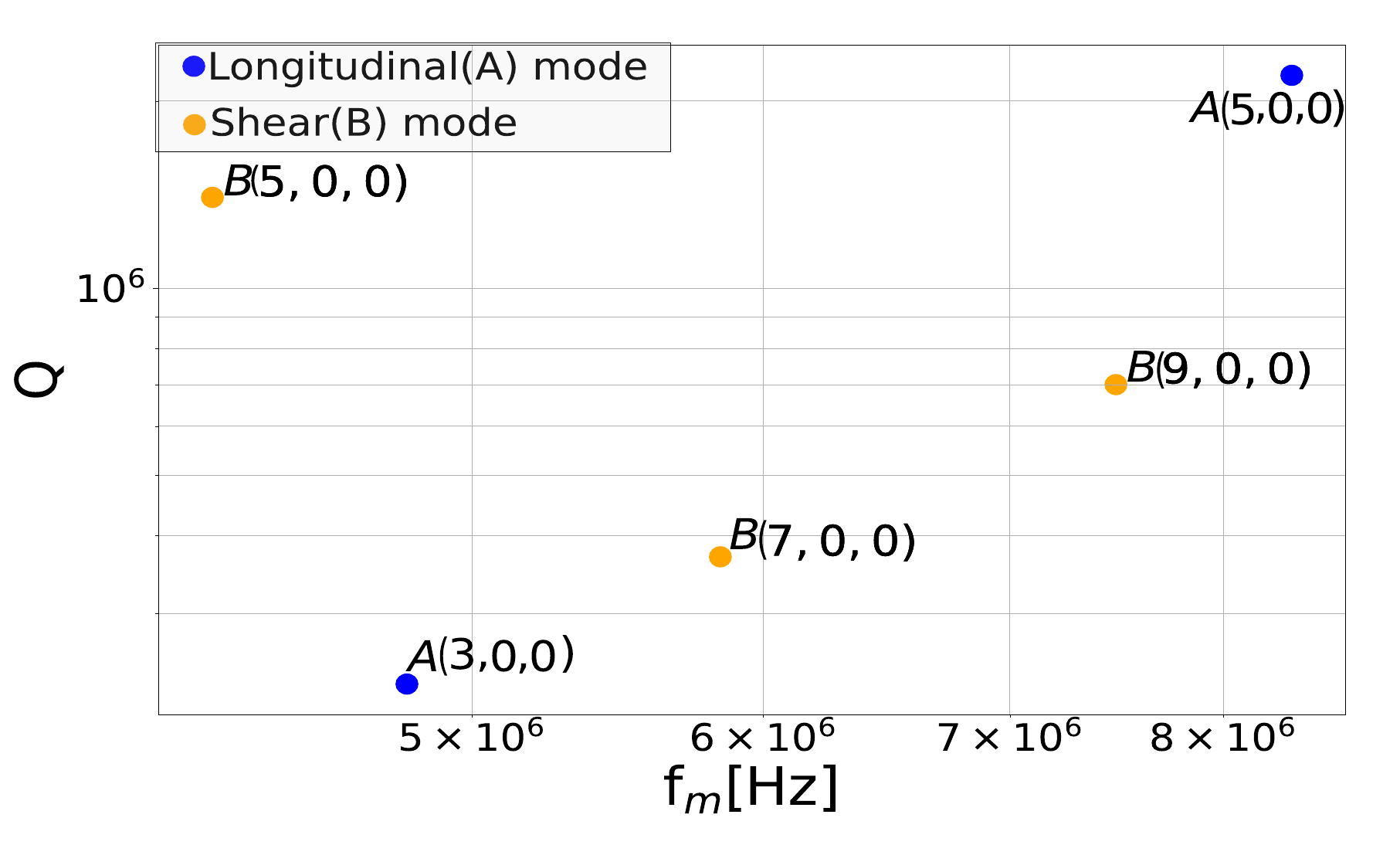}% Here is how to import EPS art
    \caption{$Q$ values for the A$_{3,0,0}$ and A$_{5,0,0}$ longitudinal and B$_{5,0,0}$, B$_{7,0,0}$, and B$_{9,0,0}$ shear modes, obtained from mixer trace output. The $Q$-factors demonstrate insignificant power dependence until the onset of duffing nonlinearity. The $Q\times f$ product of the best modes are of order $10^{14}$}
    \label{$Q$-value estimation by OME}
    \end{figure}

Fig. \ref{fig:8.35MHz-5.83MHz} c) and d) shows the mixer output voltages for the longitudinal A$_{5,0,0}$ mode and the shear B$_{7,0,0}$ modes respectively. Normalised reflection coefficients S$_{11}$ are measured by driving the BAW through the anodised post and measuring the reflected power on an impedance analyser, which is shown as the green trace. Fig. \ref{fig:8.35MHz-5.83MHz} illustrates that the mixer output voltage is indeed a result of the mechanical mode being driven.
We can determine the linewidth of the mechanical modes from the mixer output voltage and obtain $Q$-factors. These values closely agree with the previously measured $Q$ at 4 K using a parallel cathode plate \cite{campbell2024low}. Fig. \ref{$Q$-value estimation by OME} shows measured mechanical $Q$-factors at 4 K for various A and B modes.

The effective mode mass for various A and B modes of the BAW vary slightly from each other, depending on the value of their lumped circuit parameters. These values were calculated using COMSOL finite element modelling, similar to the analysis for quartz, as detailed in the works of ~\cite{bourhill2020generation, bushev2019testing}, and related analytical effective mass calculations by \cite{stevens1986analysis,shi2014variational}, with details given in Appendix \ref{sec:A1}. Various determined parameters for each of the mechanical modes are presented in Table \ref{tab:LiNbO3 coupling table}, where co-operativity $C_0$ is defined as $C_0 = 4 g^2_0/\Gamma_m \kappa_c$, $\Gamma_m$ is the line width of the mechanical resonance and $\kappa_c$ is the linewidth of the microwave cavity \cite{aspelmeyer2014cavity}.  As observed, different modes show different coupling rates due to different displacement polarisations and different overlaps between the re-entrant post-mode and the mechanical modes. 
            \begin{table}[b]
            \centering
            	\begin{tabular}{llllll}
            		\hline
            		\hline
            		$X_{n,m,p}$ & $f_m$ & $M_{\text{eff,m}}$ &  $g_0$ & $C_0$ &\\
            		\hline
            		\hline
            		B$_{5,0,0}$ & 4.20 & 3.53 $\times10^{-4}$ & 2.67 $\times10^{-7}$   &2.40$\times10^{-20}$&\\
            		A$_{3,0,0}$ & 4.70 & 4.15 $\times10^{-4}$ & 1.44 $\times10^{-5}$   &2.40$\times10^{-17}$&\\
            		B$_{7,0,0}$ & 5.80 & 3.14 $\times10^{-4}$ & 2.12 $\times10^{-8}$   &2.96$\times10^{-23}$&\\
            		B$_{9,0,0}$ & 7.40 & 2.77 $\times10^{-4}$ & 4.79 $\times10^{-8}$  &7.62$\times10^{-22}$&\\
            		A$_{5,0,0}$ & 8.30 & 3.52 $\times10^{-4}$ & 2.38 $\times10^{-6}$  &1.96$\times10^{-18}$&\\
            		\hline
            		\hline
            	\end{tabular}
            	\caption{\label{tab:LiNbO3 coupling table} Comparison of the parameters for some BAW modes. Frequencies, $f_m$, in MHz\cite{campbell2024low}; $M_{\text{eff,m}}$ in gram, evaluated from finite element modelling; $g_0$ the coupling rate in Hz, and the single photon cooperativity, $C_0$.}  
            \end{table}

\begin{figure}[t]
\centering
\includegraphics[width=1.0\linewidth]{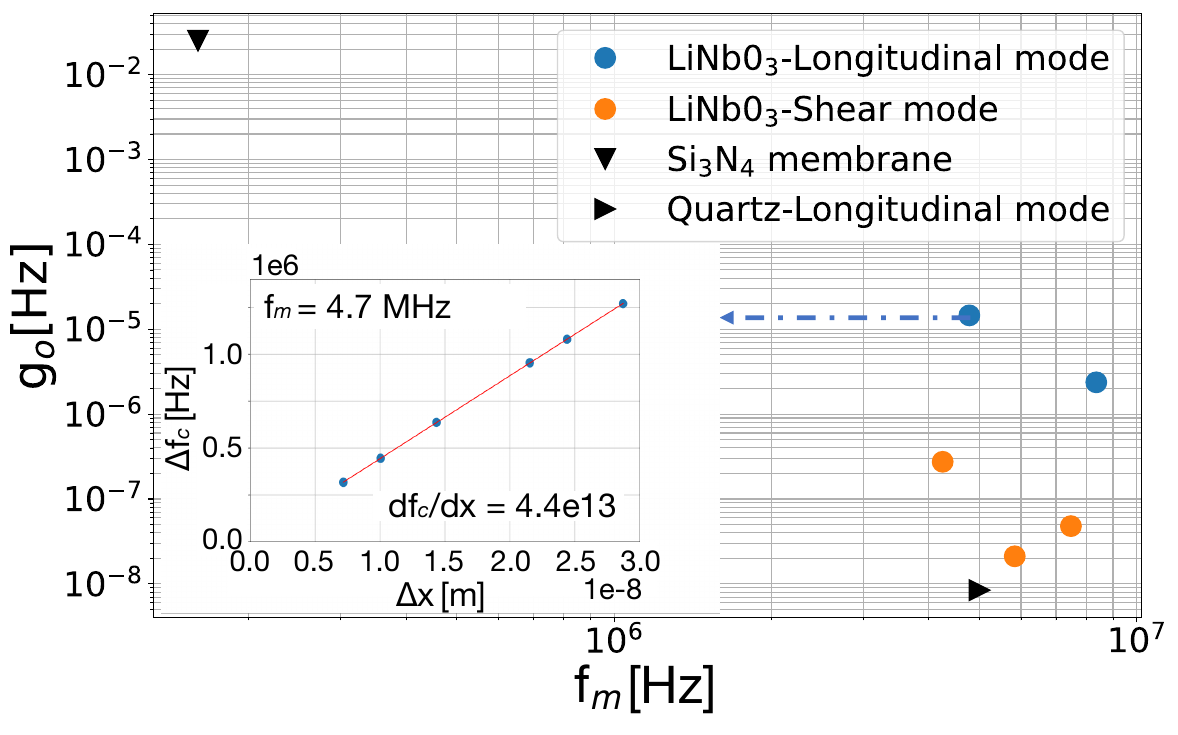}% Here is how to import EPS art
\caption{\label{Coupling rates for different experiments} Coupling rates for various A (blue) and B (orange) modes of the LiNbO$_3$-BAW at 4K, compared to other acoustic resonator coupling rates  \citep{bourhill2020generation, carvalho2019piezo,kumar2024optomechanically}. The inset figure shows the change in microwave frequency $\Delta$f$_c$ (Hz) corresponding to displacement $\Delta$x for A$_{5,0,0}$ mode, the slope of which multiplied by x$_{\text{zpf}}$ gives $g_0$.}
\end{figure}

The displacement of the crystal structure is calculated from the electrical current arising from the applied piezoelectric voltage using equations \eqref{4} and \eqref{5}. The frequency shift from the phase bridge measurement divided by the calculated piezoelectric induced displacement provides the frequency pull factor, as detailed by equations \eqref{2} and \eqref{3}. By measuring $\Delta f_c$ for different input voltages to the piezoelectric BAW (converted to displacements), coupling rates at 4 K for different modes were calculated from the gradients of $d f_c/d x$. Fig. \ref{Coupling rates for different experiments} shows coupling rates for A and B modes of LiNbO$_3$-BAW, compared to other mechanical resonators. The inset figure shows $d f_c/d x$ for the $A_{5,0,0}$ mode and is equal to 4.4 $\times$ 10$^{13}$ Hz/m. The coupling rate $g_0$ is derived from this value by multiplying it with $x_{\text{zpf}}$.

The value of these coupling rates is benchmarked against similar experiments using a quartz BAW at 4 K with a single post-re-entrant resonator, as described in \citep{bourhill2020generation, carvalho2019piezo}. The coupling rates $g_0$ for LiNbO$_3$-BAW in the split-post re-entrant cavity are observed to be up to four orders of magnitude greater compared to quartz for the longitudinal mode at A$_{3,0,0}$, demonstrating higher coupling rates and making it a suitable candidate for further studies.  Additionally, we compared our results with a Si$_3$N$_4$ membrane inside a re-entrant cavity, which achieved coupling rates of 25 mHz \cite{kumar2024optomechanically}. These findings motivate further investigation into the coupling of LiNbO$_3$-BAW with photonic systems.

The degree to which the electromagnetic and mechanical mode volumes overlap plays a key role in determining the optomechanical coupling. The improvement in coupling rates resulting from the use of a split-post versus a single-post cavity can be estimated by the ratios of a so-called overlap factor. This factor can be determined from the percentage of electromagnetic mode that exists within the mechanical mode volume determined by the dashed lines in Fig. \ref{fig:8.35MHz-5.83MHz}. The A$_{5,0,0}$ mechanical mode, for example, has all of its displacement field within a diameter of approximately 6.0 mm. Inside this volume, the electromagnetic field distribution for both the split- and single-post cavities is primarily determined by the $E_z$ field, which makes up $>99\%$ of the electrical energy, which is only 5$\%$ in the case of a single post. The amount of electrical field within the mechanical mode volume is 20 times larger for a split-post design compared to the single-post design, explaining one of the contributions to the observed coupling rate increase of these results compared to the results in quartz \citep{bourhill2020generation, carvalho2019piezo}. The remaining contribution stems from the material properties.

In conclusion, we investigated the acoustic modes of a LiNbO$_3$-BAW resonator by directly exciting them with a radio frequency external voltage at mechanical resonance through piezoelectricity at 4 K. Coupling rates of order 0.014 mHz were obtained for the A$_{3,0,0}$ longitudinal mode, four orders of magnitude larger than previous results obtained using quartz. Thus, with further development, could be utilised as an alternate material in quantum sensing and communication, and testing of fundamental physics. %In future work, we plan to explore improvements in coupling to anti-symmetric acoustic modes and by symmetrically positioning the BAW element, exploring quadratic optomechanical coupling in this system.

\begin{acknowledgments}
This research was supported by the ARC Centre of Excellence for Engineered Quantum Systems (EQUS, No. CE170100009) along with support from the Defence Science
and Technology Group (DSTG) as part of the EQUS Quantum Clock Flagship program. Additional support provided by the ARC Centre of Excellence for Dark Matter Particle Physics (CDM, No. CE200100008).
\end{acknowledgments}

\section*{Appendix: Some Acoustic Mode Properties}
\subsection{Effective mass of the mechanical modes}
\label{sec:A1}
Evaluation of mode mass was done using COMSOL, evaluating potential energy integral for 3D/2D model, for a given mode is given in terms of following equations. 
 \begin{equation}
         dU = \frac{1}{2}\omega_m^2 |a_m(t)|\bold{r}_m(\bold{x})^2\rho(\bold{x})dV .
 \end{equation}
 The total potential energy of the mode is then given by, integral over the entire volume as represented by
 \begin{equation}
     U = \frac{1}{2}m_{\text{eff,m}}\omega_m^2|a_m(t)|^2 .
 \end{equation}
 Here material density is uniform and effective mass for a 3D model 
  \begin{equation}
     m_{\text{eff,m}} = \rho \int |\bold{r}_m(\bold{x})|^2dV .
 \end{equation}
This process begins by evaluating the point displacement at the calculation point, as in the case of the odd mode, where the displacement at the point of maximum displacement is used to normalize the displacement function $\bold{r_m(x)}$. The integral of the normalized displacement over the entire geometry of the resonator is then multiplied by $\frac{\pi}{2}r\rho$ to obtain the effective mass (our model in COMSOL was a 2D axisymmetrical model). By investigating the potential energy U for any mode with a small volume element dV centred on 
$\bold{x}$ the potential energy centred around the mass element $dm = \rho(\bold{x})dV$ and is defined as below, where m$_{\text{eff,m}}$ is defined as effective modal mass for a given mode m.

\subsection{Power dependence of the mechanical mode}
\begin{figure}[h]
\centering
\includegraphics[width=0.8\linewidth]{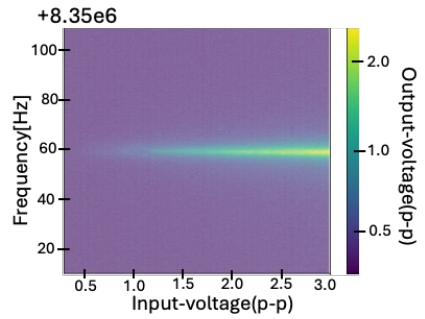}% Here is how to import EPS art
\caption{\label{Density plot for different input power} Mixer output voltage for varying input power of the A(5,0,0) mode.}
\end{figure}

Fig. \ref{Density plot for different input power} shows the density plot for A(5,0,0) mode where the input voltage to the cavity is varied in the range of 0.1 V-3.0 V, and the output of the mixer is plotted for different input power.  As the power is increased the amplitude of the sideband increases, however there is no significant change in line-width, which indicated no variation in Q-factor.

\section*{References}
%aipnum4-2.bst 2019-01-14 (MD) hand-edited version of apsrev4-1.bst
\providecommand{\noopsort}[1]{}\providecommand{\singleletter}[1]{#1}%

%aipnum4-2.bst 2019-01-14 (MD) hand-edited version of apsrev4-1.bst
\providecommand{\noopsort}[1]{}\providecommand{\singleletter}[1]{#1}%

\end{document}